\documentclass[review,12pt]{elsarticle}
\usepackage{lineno,hyperref}
\modulolinenumbers[5]

\journal{Journal of \LaTeX\ Templates}

\usepackage{amsmath}
\usepackage{graphicx}

\title{Hypersound Absorption of Acoustic Phonons in a degenerate Carbon Nanotube}

\author[rvt]{K. A. Dompreh\corref{cor1}\fnref{fn1}}
\author[focal]{N. G. Mensah}
\author[rvt]{S. Y. Mensah}
\author[rvt]{S. S. Abukari}
\author[rvt]{F. Sam}
\author[rvt]{R. Edziah}
\address[rvt]{Department of Physics, College of Agriculture and Natural Sciences, U.C.C, Ghana.}
\address[focal]{Department of Mathematics, College of Agriculture and Natural Sciences, U.C.C, Ghana}
\cortext[cor1]{Corresponding author\\ Email: kwadwo.dompreh@ucc.edu.gh}
\ead[url]{kwadwo.dompreh@ucc.edu.gh}

\date{}

\begin{document}
\begin{abstract}
\noindent Absorption of acoustic phonons  was studied in degenerate Carbon Nanotube (i.e. 
where the electrons are found close to the Fermi level). The 
calculation of the hypersound absorption coefficient ($\Gamma$) was done  in the regime where $ql>>1$ ($q$ is the acoustic phonon number and $l $
is the electron mean free path).    At $T = 10K$ and  $\theta > 0$ ($\theta$ being scattering angle),  
the dependence of $\Gamma$ on  acoustic wave number ($\vec{q}$), frequency ($\omega_q$), and  
$\gamma = 1-\frac{V_D}{V_s}$, ($V_s$  and $V_D$ being the 
speed of sound and  the drift velocity respectively) were analysed numerically  at $n = 0, \pm 1, \pm 2$ (where $n$ is an integer ) and  
presented graphically. It was observed that  when 
$\gamma < 0$, the maximum amplification was attained at  $V_D = 1.1V_s$ which occurred at 
$E = 51.7Vcm^{-1}$. In the second harmonics, ($n =\pm 2$), the absorption obtained 
was compared to experimental measurement of acoustoelectric current via the Weinreich relation   
and the results qualitatively agreed with each other. 

\end{abstract}

\maketitle
\section*{Introduction}
Carbon Nanotubes (CNT) are graphite sheets seamlessly wrapped into  cylinders
characterised by a chiral index (m,n), with m and n being two
integers, which specify the carbon nanotube uniquely~\cite{1,2,3}. This has recently 
attracted lot of interest for use in many semiconductor devices
due to its remarkable electrical~\cite{4}, mechanical~\cite{5}, and  thermal~\cite{6,7,8,9} properties 
which are mainly attributed to its unusual band structures~\cite{10}. The $\pi$-bonding and anti-bonding ($\pi^*$) energy band of CNT crosses at the 
Fermi level in a linear manner~\cite{10}.  
In the linear regime, electron-phonon interactions in CNT at low 
temperatures leads to the emission of large number of coherent acoustic phonons ~\cite{6,7,8}. 
Studies of the effect of phonons  on thermal transport~\cite{9,10}, on Raman scattering~\cite{11}
and on electrical transport~\cite{12} in CNT is an active area of research. Also, the speed of electrons in 
the linear region is extremely high. This makes CNT
a good candidate for application of high frequency electronic systems such as field effect transistors 
(FET's)~\cite{13}, single electron memories~\cite{14} and chemical sensors~\cite{15}. Another important 
investigation in  the linear regime is interaction of acoustic phonons  with  drift charges in CNT.
It is well known that when acoustic phonons interact with charge carriers, it is accompanied by  
energy and momentum exchange which give rise to the following  effects: Absorption (Amplification) 
of acoustic phonons~\cite{16,17}; Acoustoelectric Effect (AE)~\cite{18, 19, 20, 21, 22, 23}; Acoustomagnetoelectric Effect
(AME)~\cite{24,25, 26,27, 28}; Acoustothermal Effect~\cite{29} and Acoustomagnetothermal Effect~\cite{29}.
The idea of acoustic wave amplification in bulk material was theoretically predicted by Tolpygo (1956), Uritskii~\cite{30},
and  Weinreich~\cite{31} and in N-Ge by Pomerantz~\cite{32}. In Superlattices, the effect of hypersound  
absorption/amplification was extensively studied by Mensah et. al~\cite{33,34,35,36,37},Vyazovsky et. al.~\cite{38}, 
Bau et. al~\cite{39}, while Shmelev and Zung~\cite{40} calculated the absorption coefficient and renormalization 
of the short-wave sound velocity. Azizyan ~\cite{41} calculated the absorption coefficient in a quantized electric field.  
Furthermore, Acoustic wave absorption/amplification in Graphenes~\cite{42,43,44}, Cylindrical quantum wires~\cite{45} , 
and quantum dots~\cite{46,47} have all received attention.  On the concept of Acoustoelectric effect (AE) in bulk ~\cite{ 48} 
and low-dimensional materials~\cite{49}, much research has been comprehensively
done both theoretically and experimentally. Acoustoelectric effect in CNT's is now 
receiving attention with few experimental work done on it. Ebbecke et. al.~\cite{50}  studied the AE current transport in a 
single walled CNT, whilst Reulet et. al~\cite{51} studied AE in CNT. But in all these research there is no theoretically studies 
of AE in CNT.  In this paper, the absorption (amplification) of hypersound in CNT in the regime $ql>>1$
($q$ is the acoustic wave number and $l$ is the electron mean free path) is considered where the acoustic wave is 
considered as a flow of monochromatic phonons of frequency ($\omega_q$). 

It is worthy to note that the mechanism of absorption (amplification) is due to Cerenkov effect.
For practical use of the Cerenkov acoustic-phonon emission, the material must have high drift velocities and large densities of electrons~\cite{17}. 
Carbon Nanotubes (CNT) has  electron mobility of $10^5$ cm$^2$/Vs at room temperature. At low temperatures 
($T = 10K$), CNT exhibit good AE effect, which indicates that Cerenkov emission can take place in it~\cite{52}. 
The paper is organised as follows: In section $2$, the kinetic theory
based on the linear approximation for the phonon distribution function is setup, where, the rate of growth of the 
phonon distribution is deduced  and the absorption coefficient ($\Gamma$) is obtained. 
In section $3$,  the final equation is analysed numerically in a graphical form  at various 
harmonics where the absorption obtained are related to the acoustoelectric current via the Wienrich relation~\cite{31}. 
Lastly the  conclusion is presented in section $4$.

\section*{Theory}
We will proceed following the works of ~\cite{48,49} where the kinetic equation for the phonon distribution is given as 
\begin{eqnarray}
\frac{\partial N_{\vec{q}}}{\partial t} =\frac{2\pi}{\hbar}\sum_p\vert{C_{\vec{q}}}\vert^2 \{[N_{\vec{q}}(t) &+& 1]f_{\vec{p}}(1-f_{\vec{p}^\prime})
\delta(\varepsilon_{\vec{p}^\prime} - \varepsilon_{\vec{p}} +\hbar\omega_{\vec{p}})\nonumber\\
&-& N_{\vec{q}}(t) f_{\vec{p}^\prime}(1-f_{\vec{p}})\delta(\varepsilon_{\vec{p}^\prime} - \varepsilon_{\vec{p}} +\hbar\omega_{\vec{q}})\}\label{Eq_1}
\end{eqnarray}
where $N_{\vec{q}}(t)$ represent the number of phonons with  wave vector $\vec{q}$ at time $t$. The factor $N_{\vec{q}} + 1$ accounts for the 
presence of $N_{\vec{q}}$ phonons in the system when the additional phonon is emitted. The $f_{\vec{p}}(1-f_{\vec{p}})$ represent the probability that 
the initial $\vec{p}$ state is occupied and the final electron state $\vec{p}^\prime$ is empty whilst the factor $ N_{\vec{q}} f_{\vec{p}^\prime}(1-f_{\vec{p}})$
is that of the boson and fermion statistics.  The unperturbed electron distribution function is 
given by the shifted Fermi-Dirac function as 
\begin{equation}
f_{\vec{p}} = [exp(-\beta(\varepsilon(\vec{p}- m v_D)-\mu))]^{-1}\label{Eq_2}
\end{equation}
where $f_{\vec{p}}$ is the Fermi-Dirac equilibrium function, with $\mu$ being the chemical potential, $\vec{p}$ is momentum 
of the electron, $\beta = 1/kT$, $k$ is the Boltzmann constant and $V_D$ is the net drift velocity relative to the ion 
lattice site. In a more convenient form, Eqn($1$)
can be written as 
\begin{eqnarray}
\frac{\partial N_{\vec{q}}(t)}{\partial t} = 2\pi\vert{C_{\vec{q}}}\vert^2[{\frac{N_{\vec{q}}(t) + 1}{1 - exp(-\beta(\hbar\omega_{\vec{q}} - \hbar\vec{q}\cdot V_D))}+
{\frac{N_{\vec{q}}}{1 - exp(-\beta(\hbar\omega_{\vec{q}} - \hbar\vec{q}\cdot V_D))}}}]\nonumber\\
\times \sum_{\vec{p}}{(f_{\vec{p}} - f_{\vec{p}^\prime})\delta(\varepsilon_{\vec{p}^\prime} -\varepsilon_{\vec{p}} +\hbar\omega_{\vec{q}})}\label{Eq_3}
\end{eqnarray}
To simplifiy Eqn.($3$), the following  were utilised
\begin{equation} 
Q =\sum_{\vec{p}}{\frac{f_{\vec{p}} - f_{\vec{p}^\prime}}{\varepsilon_{\vec{p}} -\varepsilon_{p^\prime} 
-\hbar\omega_q -i\delta}}\label{Eq_4}
\end{equation}
\begin{equation}
f_{\vec{p}} =[exp(-\beta(\varepsilon_{\vec{p}} -\mu)) + 1 ]^{-1} \label{Eq_5}
\end{equation}
Given that 
\begin{equation}
\Gamma_{\vec{q}}  = -2\vert{C_{\vec{q}}}\vert^2 Im Q(\hbar\vec{q},\hbar\omega_{\vec{q}} -\hbar\vec{q}\cdot V_D)\label{Eq_6}
\end{equation}
the phonon generation rate simplifies to 
\begin{equation} 
\Gamma_{\vec{q}} = 2\pi\vert{C_{\vec{q}}}\vert^2\sum_{\vec{p}}{(f_{\vec{p}} - f_{\vec{p}^\prime})\delta(\varepsilon_{\vec{p}} 
-\varepsilon_{\vec{p}^\prime} -(\hbar\omega_{\vec{q}} - {\hbar\vec{q}}\cdot V_D))} \label{Eq_7}
\end{equation}
In Eqn.($7$), $f_{\vec{p}} > f_{\vec{p}\prime}$ if $ \varepsilon_{\vec{p}} < \varepsilon_{\vec{p}^\prime}$. When 
$\hbar\omega_{\vec{q}} - \hbar\vec{q} \cdot V_D > 0$, the system would return to its equilibrium configuration when perturbed where 
$$ N_{\vec{q}}^0 = [exp(-\beta(\hbar\omega_{\vec{q}} -\hbar\vec{q}\cdot V_D)-1)]^{-1}$$ 
But  $\hbar\omega_{\vec{q}} - \hbar\vec{q} \cdot V_D < 0$ leads to the 
Cerenkov condition of phonon instability (amplification).  The linear energy dispersion $\varepsilon(\vec{p})$ relation 
for the CNT is given as~\cite{53}
\begin{equation} 
\varepsilon(\vec{p}) = \varepsilon_0 \pm \frac{\sqrt{3}}{2\hbar}\gamma_0 b(\vec{p} - \vec{p}_0) \label{Eq_8}
\end{equation}
The $\varepsilon_0$ is the electron energy in the Brillouin zone at momentum $p_0$, $b$ is the 
lattice constant , $\gamma_0$ is the tight binding 
overlap integral ($\gamma_0 = 2.54$eV). The $\pm$ sign indicates that in the vicinity of the tangent point,
the bands exhibit mirror symmetry with respect to each point. The phonon and the electric field 
are directed along the CNT axis therefore $\vec{p}^\prime = (\vec{p}+\hbar \vec{q})cos(\theta)$.  Where 
$\theta$ is the scattering angle. At low temperature, the $kT << 1$,  Eqn.($5$) reduces to
\begin{equation} 
f_{\vec{p}} = exp(-\beta(\varepsilon(p)-\mu)) \label{Eq_9}
\end{equation}
Inserting Eqn.($8$ and $9$) into  Eqn.(7), and after some cumbersome calculations yield
\begin{equation}
\Gamma = \frac{4\hbar\pi\vert C_{\vec{q}}\vert^2exp(-\beta(\varepsilon_0 -\chi \vec{p}_0))}
{\gamma_0 b\sqrt{3}(1-cos(\theta))} \{exp(-\beta\chi(\eta +\hbar \vec{q})cos(\theta))- exp(-\beta\chi\eta)\}\label{Eq_10}
\end{equation}
where $\chi = {\sqrt{3}\gamma_0 b}/{2\hbar}$, and 
$$\eta =\frac{2{\hbar}^2\omega_{\vec{q}}(1 -\frac{V_D}{V_s}) + \gamma_0 b\sqrt{3}\hbar \vec{q} cos(\theta)}{\gamma_0 b\sqrt{3}(1-cos(\theta))}$$ 

\section*{Numerical analysis}
Considering the finite electron concentration, the matrix element can be modified as 
\begin{equation} 
\vert C_{\vec{q}}\vert^2\rightarrow \frac{\vert C_q\vert^2}{\vert{\aleph^{(el)}(\vec{q})\vert^2}}\label{Eq_11}
\end{equation}
where $\aleph^{(el)}(\vec{q})$ is the electron permitivity~\cite{48}. However, for acoustic phonons, 
$\vert{C_{\vec{q}}}\vert = \sqrt{{\Lambda^2 \hbar \vec{q}}/{2\rho V_s}}$,
where $\Lambda $ is the deformation potential constant and $\rho$ is the density of the material.
From Eq.($10$), taking $\varepsilon_0 = \vec{p}_0 = 0$, the Eqn.($10$) finally reduces to  
\begin{equation}
\Gamma =\frac{\vert {\Lambda}\vert^2\hbar^3 q^2 exp(-\beta\chi\eta)}{2\pi{\hbar\omega_q}\gamma_0 b\sqrt{3}(1-cos(\theta))} 
\{\sum_{n=-\infty}^{\infty}{\frac{exp(-n(\theta+\beta\chi\eta))}{I_n(\beta\chi(\eta +\hbar \vec{q}))} - 1}\} \label{Eq_12}
\end{equation}
where $I_n(x)$ is the modified Bessel function. The parameters used in the numerical evaluation of Eqn($12$) are: $\vert {\Lambda}\vert = 9$eV, $b = 1.42$nm,
$q = 10^7$ cm$^{-1}$, $\omega_q = 10^{12}$s$^{-1}$,$V_s = 4.7\times10^5$ cm s$^{-1}$, $T = 10K$, and $\theta > 0$. 
The dependence of the absorption coefficient ($\Gamma$) on the acoustic wave number ($\vec{q}$), the frequency ($\omega_q$)
and ($\gamma$) at various harmonics ($n = 0,\pm 1,\pm 2$) are presented below.
For $n = 0$, the graph of $\Gamma$ versus $\vec{q}$ at varying frequencies  
and that of $\Gamma$ versus $\omega_q$ for various acoustic wave numbers are shown in 
Figure $1$($a$ and $b$). In Figure $1a$, an amplification curve was observed, where the minimum value increases by increasing 
$\omega_q$ but above $\omega_q = 1.6\times10^{12} s^{-1}$, an absorption was obtained. In Figure $1b$, it was observed that 
absorption switched over to amplification when the $\vec{q}$ values were increased. 
\begin{figure}
\includegraphics[width =12cm]{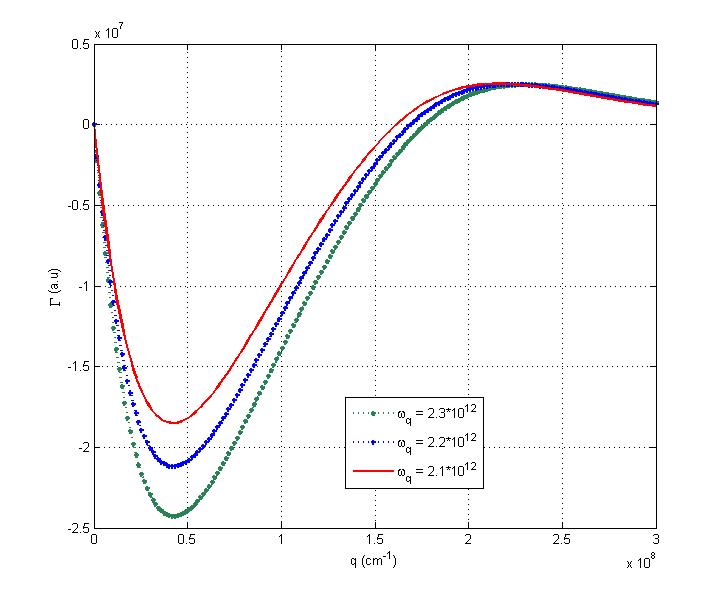}
\includegraphics[width =12cm]{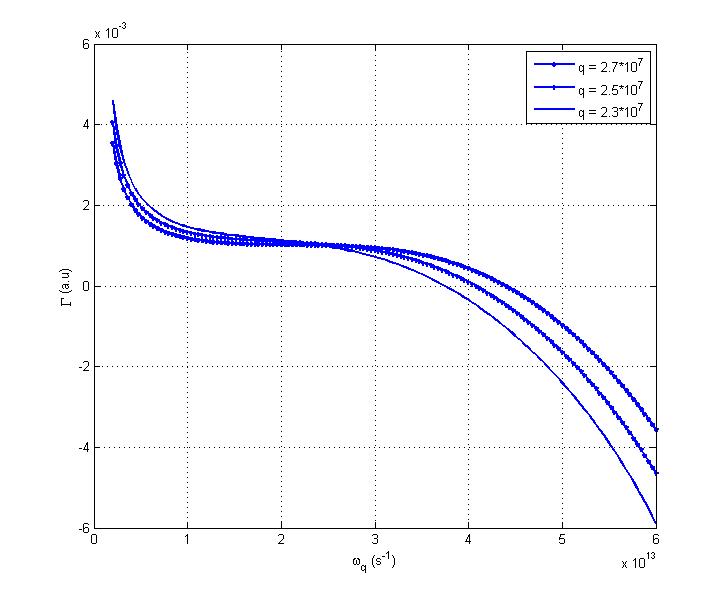}\\
\caption{(a) Dependence of $\Gamma$ on $q$  for varying $\omega_q$ at $V_D = 1.2V_s$
(b)  $\Gamma$ on $\omega_q$  for varying $\vec{q}$ at $V_D = 1.2V_s$.} 
\end{figure}
For $n =\pm 1$ (first harmonics), in Figure $2a$, it was observed  that absorption  exceed
amplification and the peaks shift to the right. A further increase in $\omega_q$  values caused
an inversion of the graph where amplification exceeds absorption (see Figure $2b$). 
A similar observation was seen in Figure $3$ ($a$ and $b$),
where, the peak values shift to the right and decreases with increasing $\vec{q}$ values 
(see Figure $3a$) but in Figure $3b$, an inversion 
of the graph occurred for increasing values of  $\vec{q}$. 
\begin{figure}
\includegraphics[width =12cm]{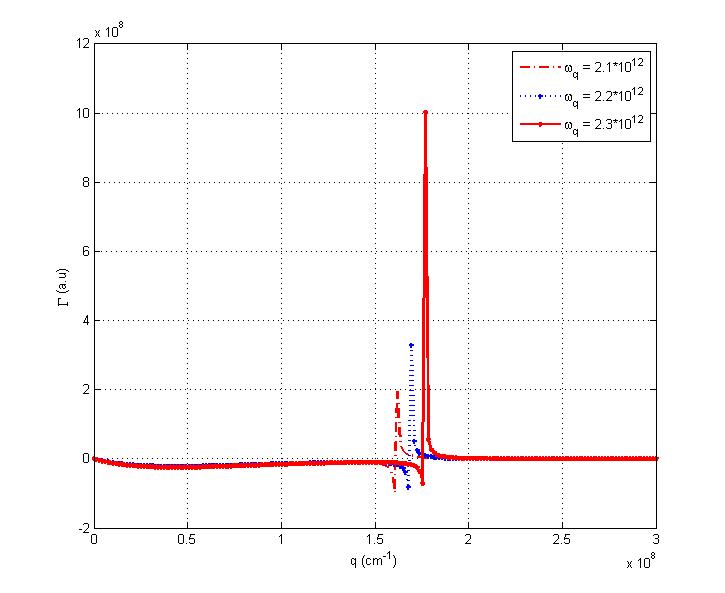}
\includegraphics[width =12cm]{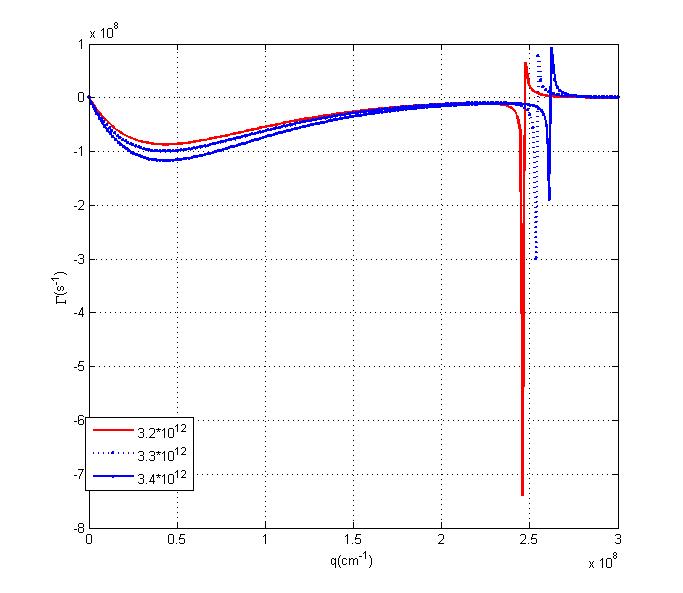}\\
\caption{Dependence of $\Gamma$ on $\vec{q}$ at $V_D = 1.2V_s$ showing (a) Absorption exceeds 
Amplification   (b) Amplification exceeds Absorption.} 
\end{figure}
\begin{figure}
\includegraphics[width =12cm]{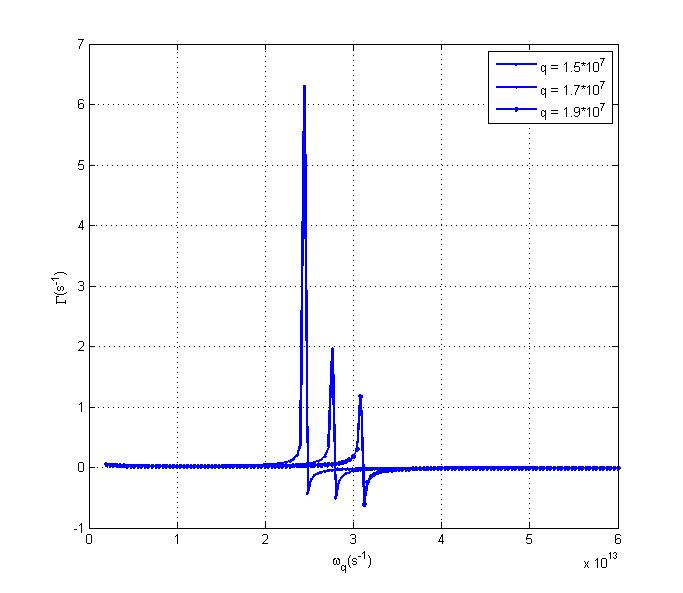}
\includegraphics[width =12cm]{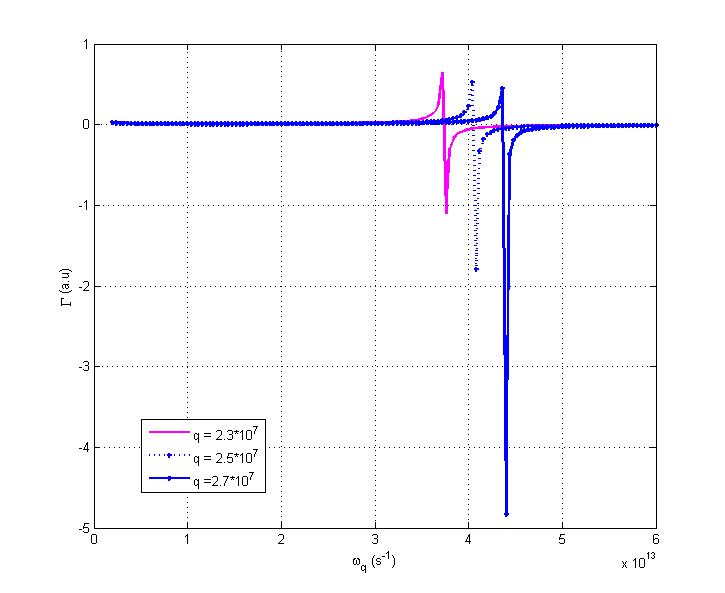}\\
\caption{ Dependence of $\Gamma$ on $\omega_q$ (a)Absorption exceeds Amplification   
(b) Amplification exceeds Absorption.} 
\end{figure}
Figure $4$ ($a$ and $b$), shows the dependence of $\Gamma$ on $\gamma$ by varying either
$\omega_q$ or $\vec{q}$. In both graphs, when
$\gamma < 0$, produce  non-linear graphs which satisfy the Cerenkov condition,
but at $\gamma > 0$, the graph returns to zero. The observed peaks 
in Figure $4a$, shift to the left by increasing $\omega_q$ whilst in Figure $4b$, 
shift to the right by increasing $\vec{q}$. 
\begin{figure}
\includegraphics[width =12cm]{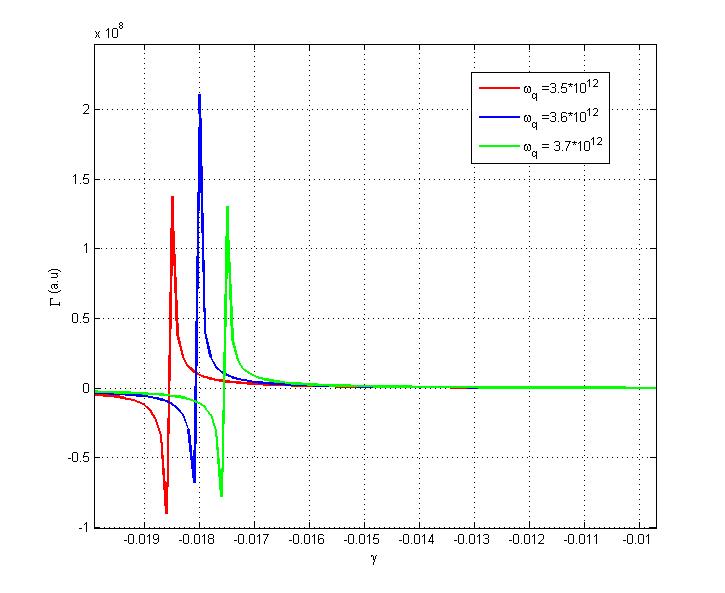}
\includegraphics[width =12cm]{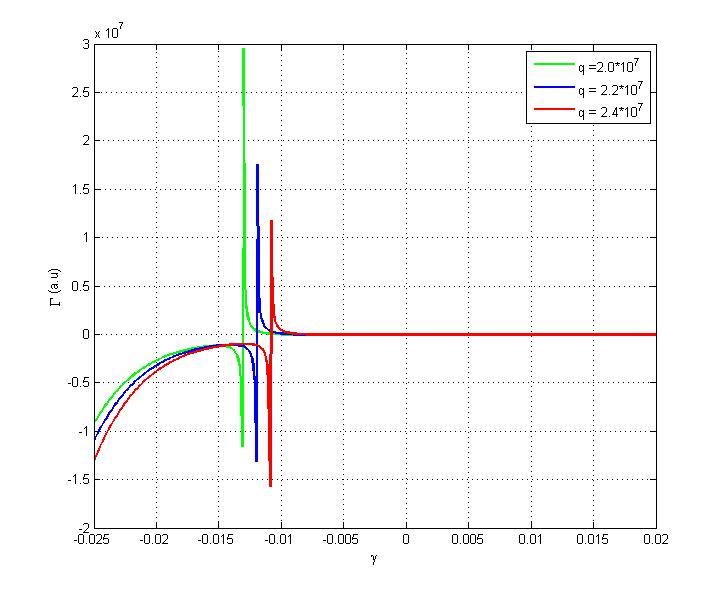}
\caption{  Dependence of  $\Gamma$ on $\gamma$ at $\theta = 80$  (a) By increasing $\omega_a$ 
(b)By increasing $\vec{q}$}
\end{figure}
For further elucidation, a $3D$ graph of $\Gamma$ versus $\omega_q$ and $\gamma$  or $\Gamma$ versus $\vec{q}$ and $\gamma$ 
are presented in Figure $5$ ($a$ and $b$). In both graphs, when $\gamma =-0.1 0$, a maximum amplification was obtained.
\begin{figure}
\includegraphics[width =12cm]{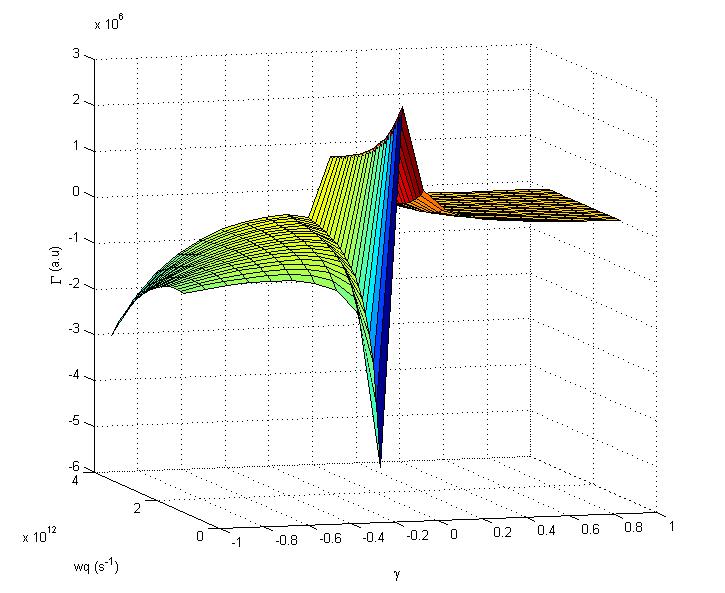}
\includegraphics[width =12cm]{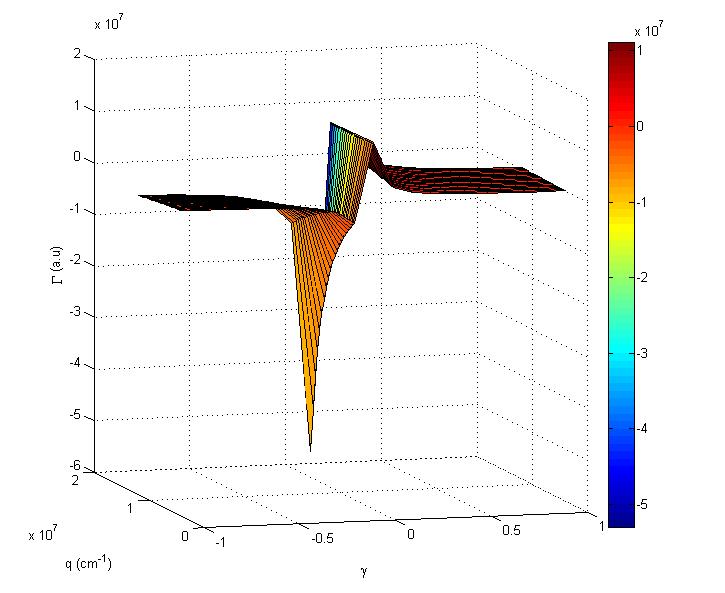}\\
\caption{(a) Dependence of $\Gamma$ on $\omega_q$ and $\gamma$ 
 and (b) Dependence of $\Gamma$ on  $\vec{q}$  and $\gamma$.} 
\end{figure}
For $n =\pm 2$ (Second harmonics), the dependence of the absorption coefficient $\Gamma$  on $\omega_q$ is 
presented in $2D$ and $3D$ form as shown in Figure $6$ and $7$. In Figure $6$, an absorption graph was 
obtained. The insert shown is an experimental results obtained for the Acoustoelectric  current in Single walled 
Carbon Nanotube~\cite{50}.  Figure $7$ ($a$ and $b$)  is the $3D$ representation of the absorption in second 
harmonics.
\begin{figure}
\center
\includegraphics[width =12cm]{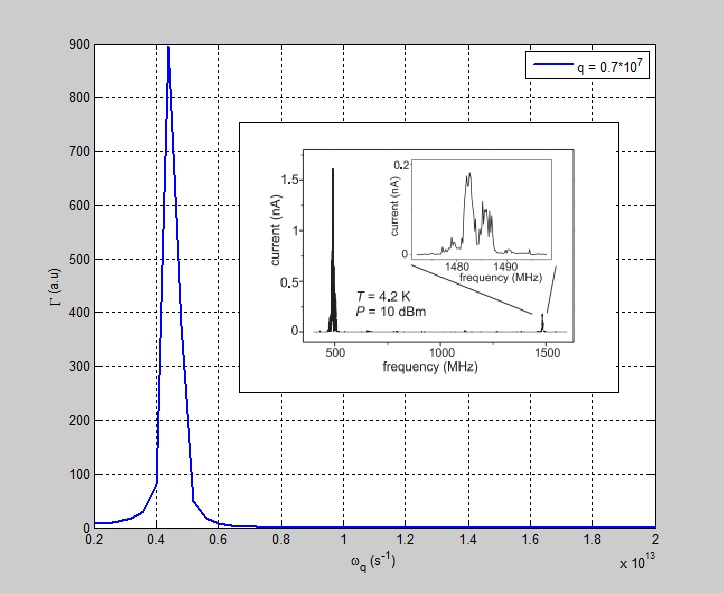}
\caption{ Second harmonic graph of the dependence of $\Gamma$ on $\omega_q$. Insert shows 
the experimental graph for acoustoelectric current versus frequency~\cite{50}} 
\end{figure} 
\begin{figure}
\includegraphics[width =12cm]{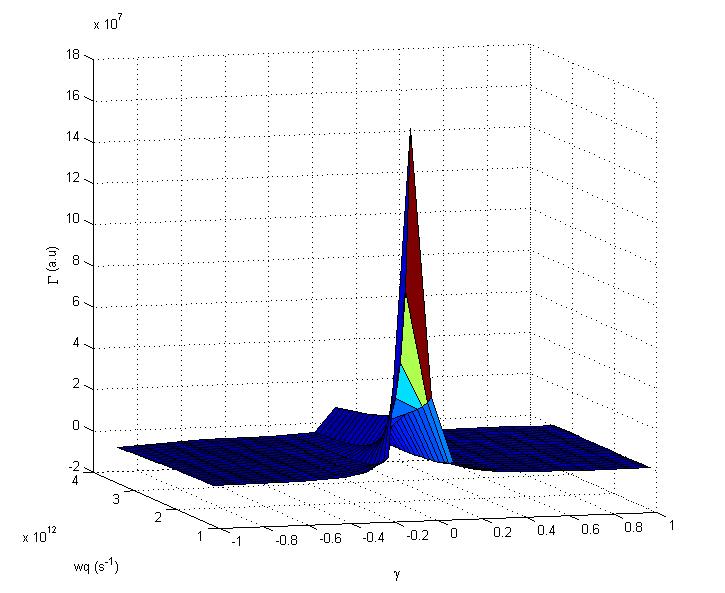}
\includegraphics[width =12cm]{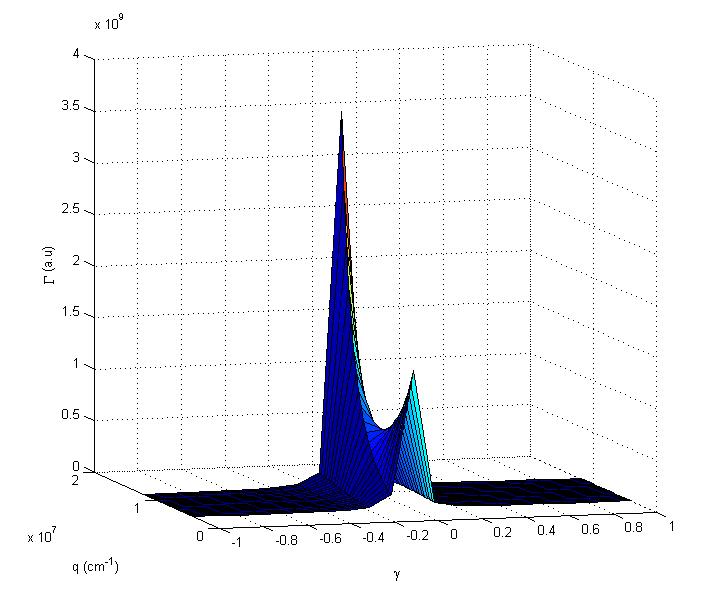}\\
\caption{(a) Dependence of $\Gamma$ on $\vec{q}$ and $\gamma$ 
 and (b) Dependence of  $\Gamma$ on $\omega_q$  and $\gamma$.} 
\end{figure}
From Weinreich relation~\cite{31}, the absorption coefficient is directly related to the acoustoelectric current, 
therefore from Figure $6$, the results obtained for the absorption coefficient qualitatively agrees with 
the experimental results presented (see insert). In the $3D$ graphs,the maximum amplification and 
absorption occurred  at $\gamma = -0.1$ which is equivalent to $V_D = 1.1V_s$. With the electric field 
$E = \frac{V_D}{\mu}$ gives $E = 51.7 V/cm$. 
\pagebreak
\section*{Conclusion}
The expression for  Hypersound Absorption of acoustic phonons in a degenerate Carbon Nanotube (CNT) was 
deduced theoretically and graphically presented.  In this work, the acoustic waves were considered to be a flow of monochromatic phonons
in the short wave region ($ql >> 1$). The general expression obtained was analysed numerically for $n = 0,\pm 1,\pm 2$
(where $n$ is an integer). From the graphs, at certain values of $\omega_q$ and  $\vec{q}$, an Amplification was
observed to exceed  Absorption or vice-versa . For $\gamma < 0$, the maximum
Amplification was observed at $V_D = 1.1V_s$ which gave us a field of $E = 51.7Vcm^{-1}$. This field
is far lower than  that observed in superlattice  and homogeneous semiconductors permitting the 
CNT  to be a suitable material for hypersound generator (SASER). A similar expression can be seen in 
the works of Nunes and Fonseca ~\cite{54}. 

Very interesting to our work is the qualitative agreement of the absorption graph to an  experimental 
graph resulting from an acoustoelectric current via the Weinriech relation.

\renewcommand\refname{Bibliography}

\end{document}